\documentclass{appolb}
\usepackage{graphicx}

\begin{document}
\title{Transitions and spin dynamics at very low temperature in the 
pyrochlores Yb$_2$Ti$_2$O$_7$ and Gd$_2$Sn$_2$O$_7$}
\author{P.Bonville, J.A.Hodges, E. Bertin, J.-Ph. Bouchaud, M. Ocio,
\address{C.E.A.- Saclay, Service de Physique de l'Etat Condens\'e
\\ 91191 Gif-sur-Yvette, France}
\and
P. Dalmas de R\'eotier, L.-P. Regnault, H. M. R\o nnow, J. P. Sanchez, 
S. Sosin, A. Yaouanc
\address{C.E.A.- Grenoble, Service de Physique
Statistique, de Magn\'etisme et Supraconductivit\'e\\
38054 Grenoble, France}
\and
M. Rams, K. Kr\'olas
\address{Institute of Physics, Jagellonian University, 30059 Krak\'ow, Poland}
}
\maketitle

\begin{abstract}
The very low temperature properties of two pyrochlore compounds, 
Yb$_2$Ti$_2$O$_7$ and Gd$_2$Sn$_2$O$_7$, were investigated using an ensemble
of microscopic and bulk techniques. In both compounds, a first order 
transition is
evidenced, as well as spin dynamics persisting down to the 20\,mK range.
The transition however has a quite different character in the two materials:
whereas that in Gd$_2$Sn$_2$O$_7$ (at 1\,K) is a magnetic transition towards
long range order, that in Yb$_2$Ti$_2$O$_7$ (at 0.24\,K) is reminiscent of the
liquid-gas transition, in the sense that it involves a 4 orders of
magnitude drop of the spin fluctuation frequency; furthermore, no long range
order is observed. These unusual features
we attribute to frustration of the antiferromagnetic exchange interaction in 
the pyrochlore lattice. 
\end{abstract}

\PACS{75.40.-s, 75.25.+z, 76.75.+i, 76.80.+y  }

\section{Introduction}

Frustration of exchange interactions among magnetic ions can arise in
different ways: the first to be studied was frustration induced by 
crystallographic disorder, leading to the so-called spin glass materials.
Frustration can also appear in fully crystallographically ordered materials, 
if the geometry of
the lattice is such that it prevents all pairs of exchange bonds from being
satisfied throughout the lattice. The simplest example is the bidimensional
triangular lattice with isotropic (Heisenberg) antiferromagnetic (AF) nearest
neighbour exchange. Villain drew attention to geometrically frustrated 
systems \cite{villain} and showed that no N\'eel order can occur
in a Heisenberg antiferromagnet on a three-dimensional lattice of corner
sharing tetrahedra.
The ground state of such a system was called a ``cooperative paramagnet'', or
a ``spin liquid'' state, where the spins undergo short range dynamic 
correlations down to $T=0$. In the past decade, investigations of 
geometrically
frustrated systems have developed to a large extent \cite{ramirez}, and a
number of
lattices have been shown to be prone to frustration: the bidimensional
\textit{kagom\'e} lattice, made of corner-sharing coplanar triangles, the
pyrochlore lattice, made of corner-sharing tetrahedra, and the garnet lattice,
made of non-coplanar corner sharing triangles. 

\begin{figure} [!ht]
\begin{center}
\includegraphics[width=0.6\textwidth]{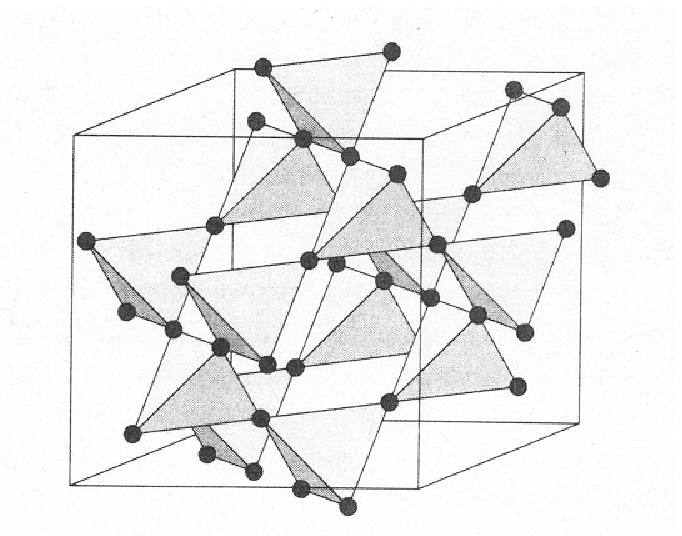}
\end{center}
\caption{\sl The pyrochlore lattice of $R_2M_2O_7$ materials, where only the 
$R$ sublattice is shown (black dots); for each $R^{3+}$ ion, the local
symmetry is threefold, with the appropriate local [111] axis as symmetry 
axis.} 
\label{stru}
\end{figure}

The formation of the ground state and the
low temperature spin dynamics have been particularly studied for the case
of the Heisenberg antiferromagnet in a pyrochlore lattice with nearest 
neighbour exchange \cite{moessner}. The ground state configuration is shown 
to have a large
degeneracy, stemming from the energy mimizing condition: 
$\sum_i \textbf{S}_i = 0$, where the sum runs over the spins $\textbf{S}_i$
of a ``plaquette'' or a ``unit'', i.e. either a triangle or a tetrahedron.
These ground states are not separated by energy barriers, and this can have
important implications as regards the spin dynamics: zero energy local or 
extended soft modes are possible, even at very low temperature. Another
consequence is that any perturbation, like the dipole-dipole interaction, next
nearest neighbour exchange, ionic or exchange anisotropy, can select a
particular ground state, resulting in a transition at a finite temperature.
The perturbations can also alter the flat energy landscape
and create energy barriers between ground configurations, which suggests
that real frustrated pyrochlore systems can retain some features of
spin-glasses.\par

Compounds with formula R$_2$M$_2$O$_7$, where R is a rare earth and M a
transition or $sp$ metal, crystallise in a structure where both R and M 
ions are located at the vertices of two interpenetrating corner sharing
tetrahedra, or pyrochlore, lattices (Fig.\ref{stru}).
In the course of our study of rare earth pyrochlores, we have discovered novel
behaviours in two materials, Yb$_2$Ti$_2$O$_7$ and Gd$_2$Sn$_2$O$_7$, which 
are the subject of this report. In these compounds, the rare earth alone is 
magnetic and it is located at a site with threefold symmetry ($D_{3d}$). 
The paramagnetic Yb$^{3+}$ ($4f^{13}$) and Gd$^{3+}$ ($4f^7$) ions
have distinct single ion properties: the crystal electric field
splits the ground spin-orbit multiplet $\{J=7/2\}$ of the Yb$^{3+}$ ion into
4 Kramers doublets, whereas it has practically no influence on the Gd$^{3+}$
ion which has $L=0$ and $S=7/2$. This implies that the
ground Yb$^{3+}$ Kramers doublet, which alone is populated at low temperature,
is described by an effective spin 1/2 with a g-tensor having uniaxial
anisotropy, whereas the Gd$^{3+}$ ion is isotropic and has a g-factor
very close to 2.\par
Our investigations were carried out
using local techniques: M\"ossbauer spectroscopy on the isotopes 
$^{170}$Yb and $^{155}$Gd, Muon Spin Relaxation ($\mu$SR) spectroscopy,
Perturbed Angular Correlations (PAC) on the isotope $^{172}$Yb and neutron
diffraction on the one 
hand, and bulk measurements: magnetic susceptibility 
and specific heat on the other hand. Most of these measurements were
performed down to the 20\,mK temperature range.\par

\section{Yb$_2$Ti$_2$O$_7$}

\begin{figure} [!ht]
\includegraphics[width=0.42\textwidth]{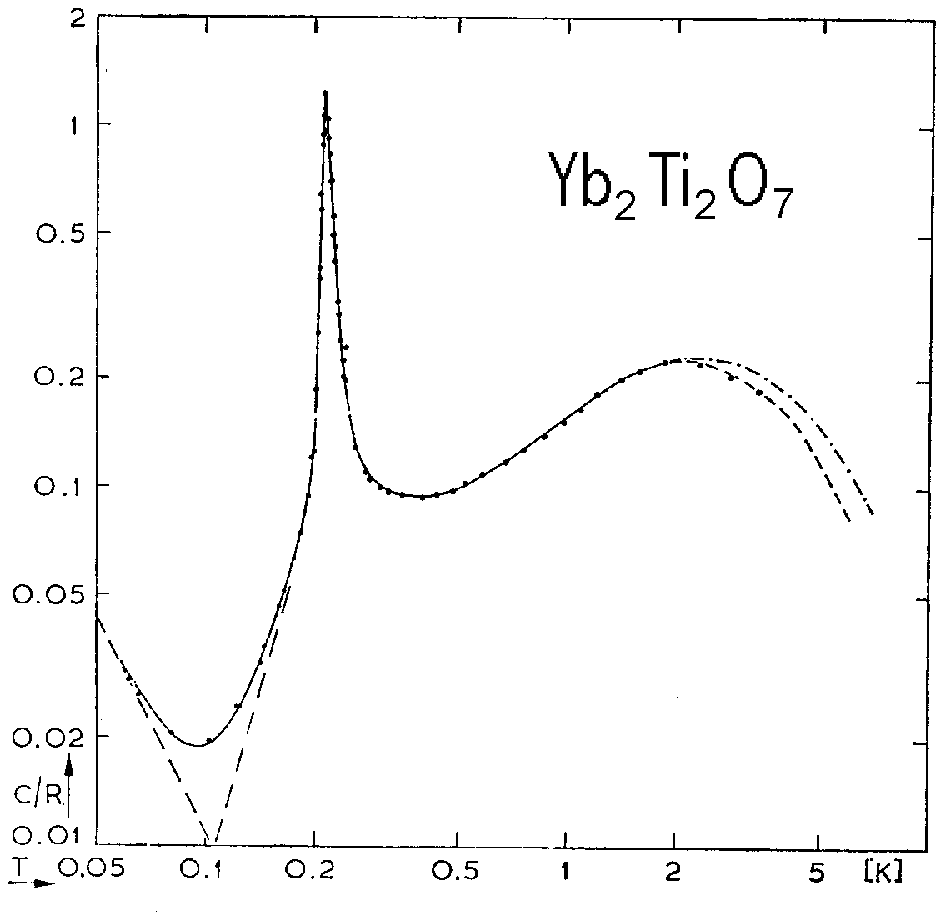}
\hspace{0.5 cm}
\includegraphics[width=0.5\textwidth]{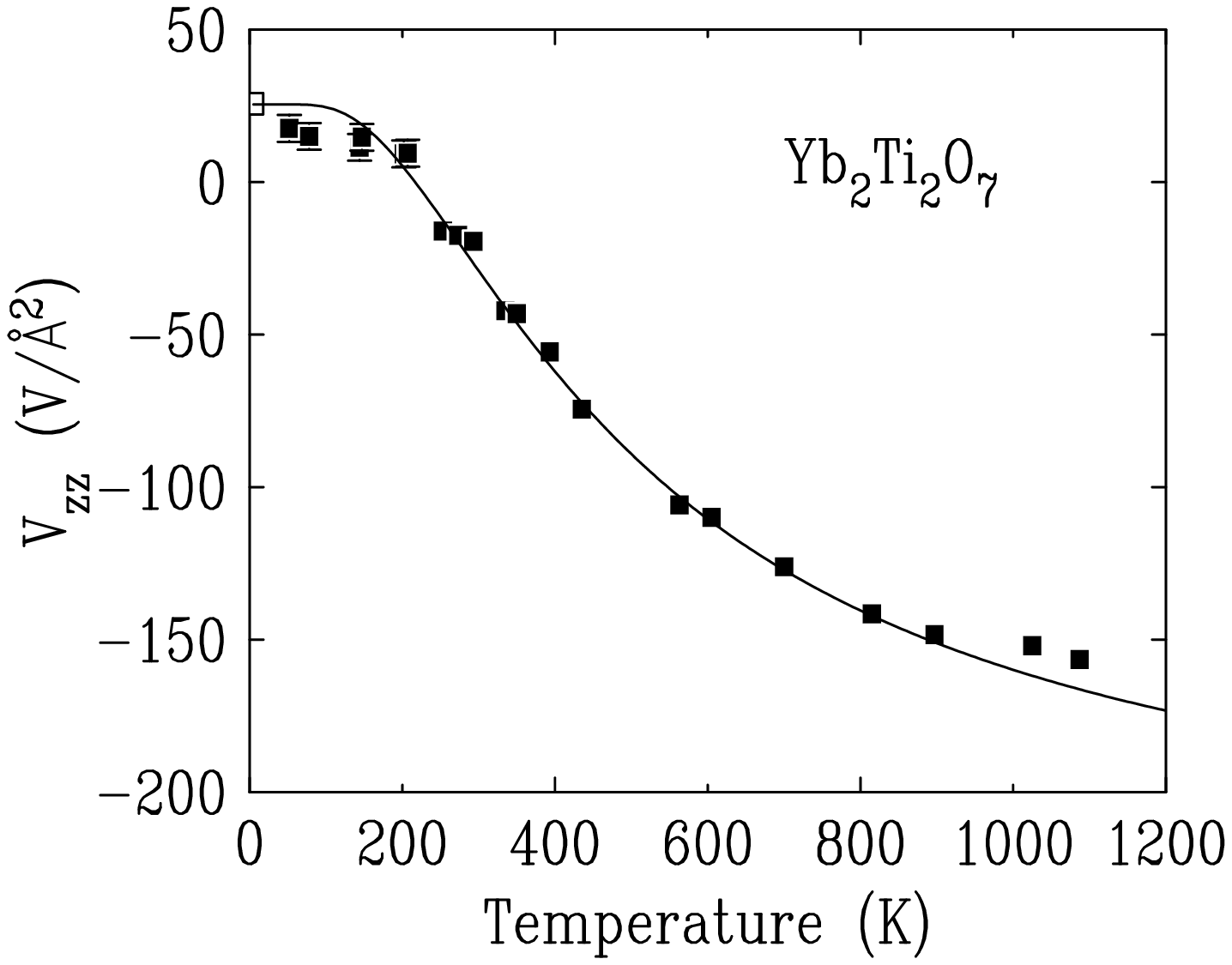}
\caption{\sl In Yb$_2$Ti$_2$O$_7$: \textbf{Left}: Specific heat (from 
Ref.\protect\cite{blote}),
and \textbf{Right}: thermal variation of the principal component $V_{zz}$ of 
the
electric field gradient at the  nucleus site as measured by PAC on $^{172}$Yb
(solid symbols) and $^{170}$Yb M\"ossbauer spectroscopy (open symbol); the
line is a fit to a crystal electric field model (see text).}
\label{cpvzz}
\end{figure}

In Yb$_2$Ti$_2$O$_7$, a sharp peak in the specific heat had been evidenced 
near 0.24\,K \cite{blote} (Fig.\ref{cpvzz} left), but the nature of the 
transition had not been further investigated. A broad feature, peaking around 
2-3\,K, is also visible. As will be shown below, the
ground crystal field doublet of the Yb$^{3+}$ ion is well isolated from the 
excited states.
Therefore, the broad feature cannot be assigned to a crystal field
Schottky anomaly, but it must be attributed to exchange spin correlations.
The entropy released at the transition amounts to only 20\% of the total
magnetic entropy associated with a doublet, $R\ln2$. This behaviour shows
that the correlations develop far above the temperature of the 
transition (0.24\,K), which is a characteristic feature of frustrated systems 
\cite{ramirez}. \par
In order to assess the crystal field splitting in this material, we performed 
PAC experiments on the isotope $^{172}$Yb, which allow
the thermal variation of the electric field gradient (EFG) tensor at the 
nucleus site to be measured. In axial symmetry, the EFG tensor is specified
by its principal component $V_{zz}={{\partial^2V} \over {\partial z^2}} 
(\textbf{r}=0)$ alone, which is the sum of two terms:
\begin{equation}
V_{zz}(T) = V_{zz}^{latt} + B_Q \langle 3J_z^2 -J(J+1) \rangle_T.
\label{vzzt}
\end{equation}
The first term is a temperature independent lattice charge contribution, and
the second term is proportional to the $4f$ shell quadrupole moment, which
is temperature dependent due to the progressive population
of the crystal field levels as temperature increases. 

The quantity $V_{zz}(T)$
decreases monotonously as temperature increases (Fig.\ref{cpvzz} right), and 
tends to $V_{zz}^{latt}$
at high temperature ($\simeq -200$\,V/\AA$^2$) because the $4f$ shell
term vanishes when all the crystal field states are equipopulated. One 
observes that $V_{zz}(0)$ is quite small, which is to a coincidental 
cancellation of the lattice and saturated $4f$ contributions, which are of 
opposite signs and of comparable magnitudes. We also performed
M\"ossbauer experiments at 4.2\,K on $^{170}$Yb diluted in non-magnetic
Y$_2$Ti$_2$O$_7$, which enabled us to measure the axially symmetric hyperfine 
tensor associated with the ground crystal field doublet. This tensor is
 proportional to the spectroscopic g-tensor, which is found to have 
components: $g_z \simeq 1.80$ and $g_\perp \simeq
4.27$. This shows that the plane perpendicular to the local [111] axis
is an easy magnetic plane for the Yb$^{3+}$ ion at low temperature. From both 
the $V_{zz}(T)$ thermal variation and the measurement of the ground state 
g-tensor (assuming that the latter is identical
for Yb in Y$_2$Ti$_2$O$_7$ and in Yb$_2$Ti$_2$O$_7$), we could determine the 
full Yb$^{3+}$ crystal 
field level scheme \cite{hodge0} in Yb$_2$Ti$_2$O$_7$. In particular, the 3 
excited doublets
lie at 620, 740 and 940\,K above the ground state. At low temperature, the
physics is therefore governed by the anisotropic ground Kramers doublet 
alone.\par

\begin{figure} [!ht]
\includegraphics[width=0.6\textwidth]{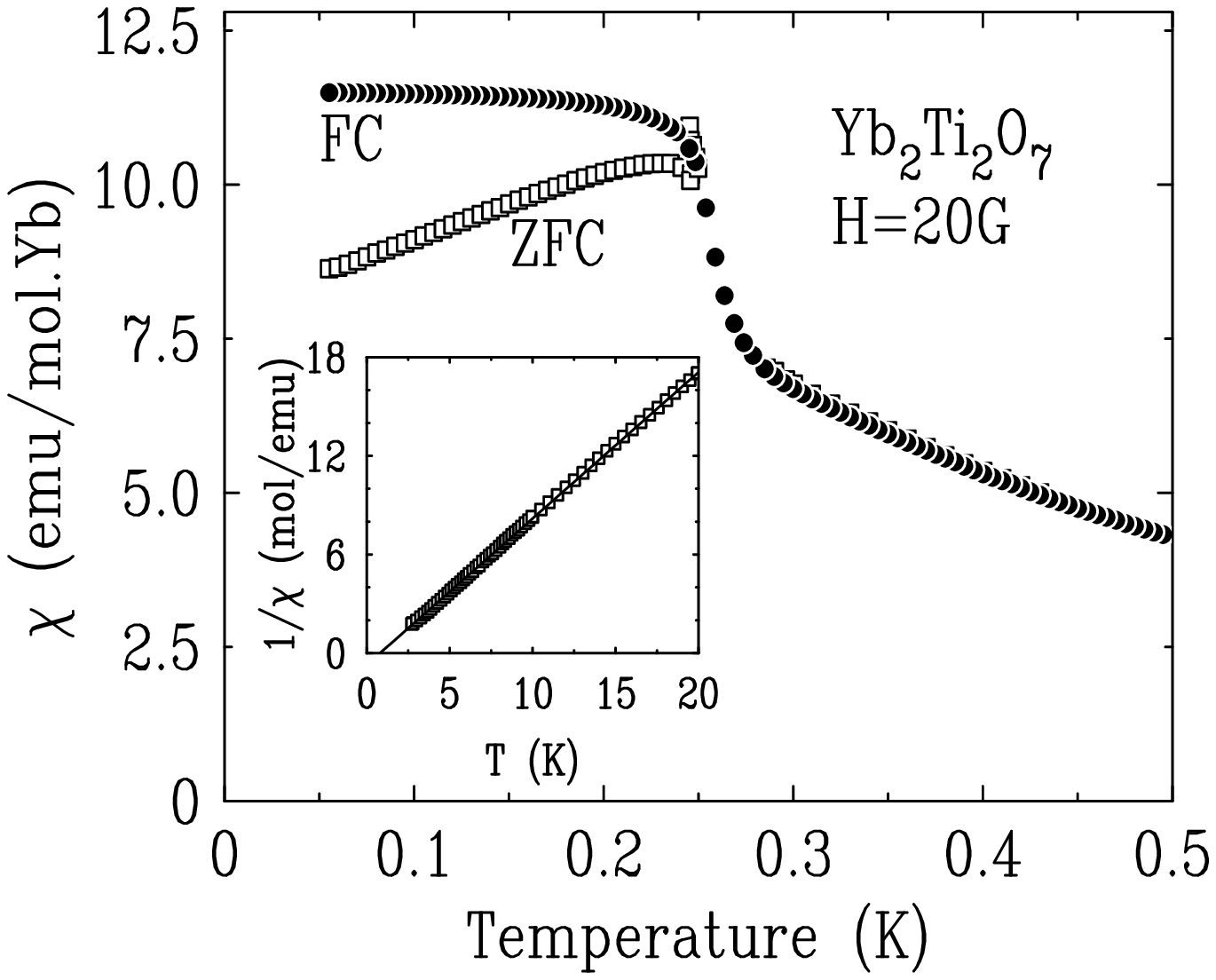}
\hspace{0.3 cm}
\includegraphics[width=0.35\textwidth]{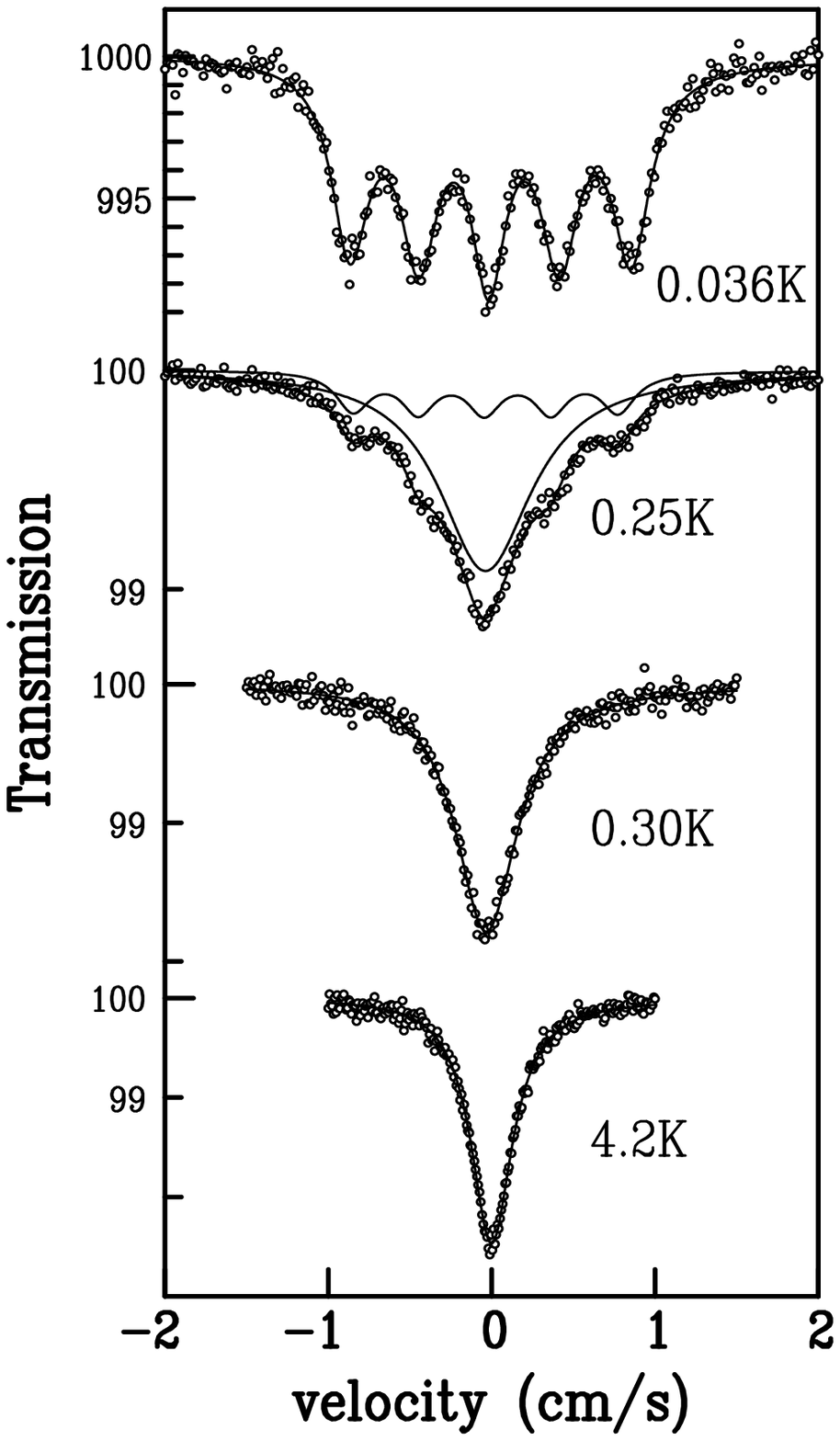}
\caption{\sl In Yb$_2$Ti$_2$O$_7$: \textbf{Left}: magnetic susceptibility
below 0.5\,K showing the two FC and ZFC branches (inset: inverse 
susceptibility up to 20\,K), and \textbf{Right}: 
selected $^{170}$Yb M\"ossbauer absorption spectra between 0.036\,K and 
4.2\,K; the lines are fits as explained in the text.}
\label{ximoss}
\end{figure}

The magnetic susceptibility $\chi(T)$ follows a Curie-Weiss law below about 
50\,K,
with a small positive paramagnetic Curie temperature $\theta_p \simeq 0.8$\,K
(inset of Fig.\ref{ximoss} left). This can be indicative of a 
ferromagnetic net exchange interaction of magnitude $ca.$ 1\,K, but it does
not preclude the presence of larger nearest and next nearest 
neighbour couplings with opposite signs. This latter assumption seems to be
supported by the neutron diffraction data on a single crystal, to be described
below. At 0.24\,K, the temperature of the
specific heat peak, an anomaly is seen in $\chi(T)$ (Fig.\ref{ximoss} left). 
Below 0.24\,K irreversibilities between the Field Cooled (FC)
and Zero Field Cooled (ZFC) branches appear, which are reminiscent of a
spin-glass behaviour. \par
Selected M\"ossbauer absorption spectra on the isotope $^{170}$Yb ($I_g$=0, 
$I_e$=2, $E_0=84.3$\,keV), represented in Fig.\ref{ximoss} right, reveal an 
apparently standard first order magnetic 
transition at 0.24\,K. At 0.036\,K, a five-line magnetic 
hyperfine 
pattern is observed with a hyperfine field $H_{hf} \simeq 115$\,T 
corresponding to a
saturated Yb$^{3+}$ moment of $ca.$ 1.15\,$\mu_B$. As temperature increases,
the spectrum remains unchanged up to 0.22\,K; then a broad single line grows
superimposed on the five-line spectrum, and at 0.26\,K the single line alone
is left. In the temperature range 0.036\,K $\le T \le$ 0.26\,K, the hyperfine 
field associated with the
five-line spectrum remains essentially constant. This fact and the 
coexistence, in a 
small temperature region, of the five-line pattern and of the single line,
is characteristic of a first order transition. In principle, for a first order
transition, one phase transforms into the other at the transition temperature
by absorbing or releasing latent heat. In real systems, however, there is a 
narrow distribution of transition temperatures among the crystallites; this
leads to the coexistence of the two phases in a small region around the mean
transition temperature.\par
For an anisotropic Kramers doublet, the saturated moment value depends on the 
orientation of the applied or exchange field and, except when the field is 
along a principal direction of the g-tensor, the moment and the field are not
collinear. The relationship linking the modulus $m$ of the saturated moment 
and the angle $\theta$ it makes with the local $z$-axis is \cite{hodge0}:
\begin{equation}
\cos^2\theta = {{({m_\perp \over m})^2 -1} \over {r^2 -1}},
\label{thetam}
\end{equation}
where $m_\perp = {1 \over 2} g_\perp \mu_B$ and $r = g_\perp / g_z$. In
Yb$_2$Ti$_2$O$_7$ below 0.24\,K, from the
measured $m$ value and known $r$ value, the angle between the moment and the
local [111] axis is:$\theta \simeq 45^\circ$.
So the Yb moment does not lie in the easy plane perpendicular to [111]; the
angle $\varphi$ the exchange field makes with the local $z$-axis can also be 
obtained through the relation: $\tan\theta = r^2 \tan \varphi$. One gets:
$\varphi \simeq 10^\circ$. So the exchange field is almost directed along the
threefold symmetry axis. At the present time, the origin of these angles 
cannot be reasonably interpretated, but they may provide clues for a future
investigation of the spin configuration in Yb$_2$Ti$_2$O$_7$.\par

\begin{figure} [!ht]
\begin{center}
\includegraphics[width=0.6\textwidth]{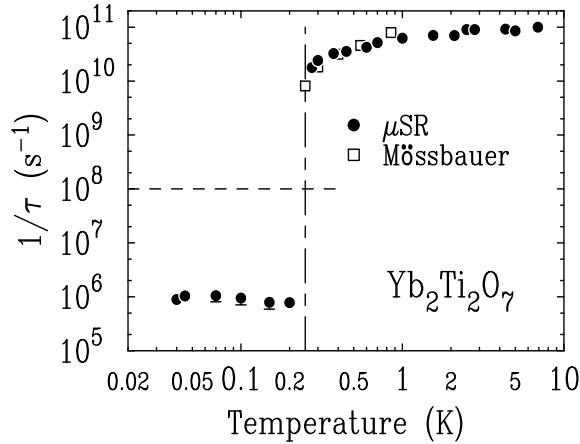}
\end{center}
\caption{\sl Thermal variation of the Yb spin fluctuation
frequency as measured by $\mu$SR (solid circles) and $^{170}$Yb M\"ossbauer
spectroscopy (open squares); the dashed line is the lower limit $\nu_l$ of the
M\"ossbauer window for measurement of fluctuation frequencies.}
\label{freqt}
\end{figure}

The characteristic time scale in these experiments is the hyperfine Larmor
period for $^{170}$Yb, which is about 10$^{-9}$\,s. This value is in fact the
center of the $^{170}$Yb ``M\"ossbauer window'' where the spin
fluctuation frequency can be measured from the lineshape \cite{gonz}. This has
two consequences in the present case. First, it
cannot be said whether the hyperfine field spectra observed below 0.24\,K
correspond to a static long range order (LRO) or to dynamic short range 
correlations with fluctuation frequencies smaller than the lower limit 
$\nu_l \simeq 10^8$\,s$^{-1}$ of the ``M\"ossbauer window''. This problem 
will be
dealt with below when describing the neutron diffraction and $\mu$SR 
experiments. Second, the spectra above 0.24\,K show a single line 
whose width decreases as temperature increases: this is the
fingerprint of the ``extreme narrowing'' regime, where the spin fluctuation
frequency is much larger than the hyperfine coupling and increases as
temperature increases. The fluctuation frequency can be measured as long as it
is below the upper limit of the ``M\"ossbauer window'' for $^{170}$Yb, i.e. 
$\nu_u \sim 5\times 10^{10}$\,s$^{-1}$. A physical hypothesis about the 
fluctuations
is also needed: for paramagnetic fluctuations, a lineshape of the type 
developed in Ref.\cite{gonz} is adequate, but for hyperfine field 
fluctuations (I.e. correlated moments), a stochastic lineshape as described 
in Ref.\cite{dattagupta}
must be used. In Yb$_2$Ti$_2$O$_7$, the neutron diffraction
measurements on a single crystal, to be described below, show that spin 
correlations are present well above 0.25\,K. So we interpret the spectra
in the temperature range 0.25\,K $\le T \le$ 0.9\,K in terms of hyperfine
field fluctuations, where \textbf{H$_{hf}$} jumps 
isotropically at random with a characteristic frequency $\nu$
\cite{dattagupta}. Then the dynamical broadening of the M\"ossbauer spectrum 
in the ``extreme narrowing'' regime is given by:
\begin{equation}
\Delta \Gamma_R = {{\nu_{hf}^2} \over \nu},
\label{nuhf}
\end{equation}
where the hyperfine frequency $\nu_{hf}$ is $\mu_I H_{hf}$, $\mu_I$ being the
magnetic moment of the excited $^{170}$Yb nuclear state. The fluctuation
frequencies obtained thereby are reported in Fig.\ref{freqt} and will be
discussed below, together with the $\mu$SR data.\par

A first answer to the question about the presence of LRO in 
Yb$_2$Ti$_2$O$_7$ comes from neutron diffraction experiments on a powder 
sample (performed at the ``Laboratoire L\'eon Brillouin'', Saclay), which 
reveal there are no magnetic Bragg peaks below 0.24\,K. Therefore there is no 
LRO in Yb$_2$Ti$_2$O$_7$ below the temperature of the specific heat peak.

\begin{figure} [!ht]
\begin{center}
\includegraphics[width=0.7\textwidth]{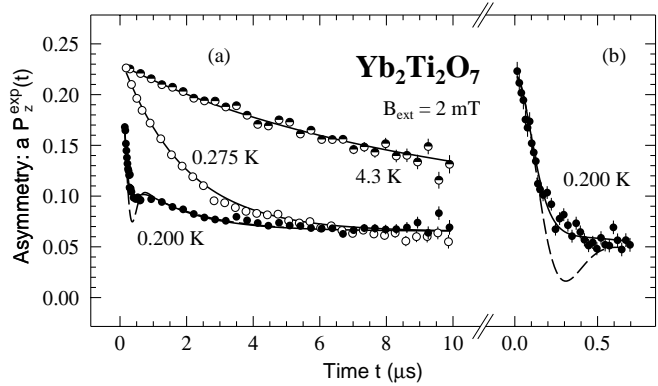}
\end{center}
\caption{\sl In Yb$_2$Ti$_2$O$_7$: (a) $\mu$SR depolarisation spectra,
measured at ISIS, on each side of the transition temperature (0.24\,K) and 
(b) detail of the depolarisation at 0.2\,K at short times, measured at PSI; 
the lines are fits as explained in the text.}
\label{musr}
\end{figure}

An insight
into the dynamics of the moments is further provided by $\mu$SR measurements
(whose characteristic time is about 10$^{-6}$\,s), performed at ISIS 
(Rutherford Appleton Laboratory, Chilton, England) and at PSI (Villigen,
Switzerland) \cite{hodgprl}. For all temperatures down to 0.275\,K 
(see Fig.\ref{musr} a),
the time decay of the muon depolarisation has an exponential form:
\begin{equation}
a P_z(t) = a_z \exp(-\lambda_z t) + a_{bg},
\label{depexp}
\end{equation}
where $a_{bg}=0.065$ is a time independent background contribution arising
from muons stopping in the Ag sample holder, and $a_z \simeq 0.17$. These 
data were recorded in a
small longitudinal magnetic field $B_{ext}=2$\,mT in order to suppress the
contribution of the nuclear moments to the depolarisation. The relaxation
rate $\lambda_z$ is approximately constant ($\simeq 0.1$\,MHz) as temperature
decreases down to a few Kelvin, then it rises rapidly.
An exponential decay for the depolarisation is usually observed when the
fluctuation frequency $\nu$ of the electronic moments is much larger than
the dipolar (and hyperfine) coupling $\Delta$ between the muon spin and
the electronic spin: $\nu \gg \Delta$ (``extreme narrowing'' limit). In this
limit and in the paramagnetic phase (i.e. when the electronic moments are
uncorrelated), the following relationship links $\lambda_z$ and $\nu$:
\begin{equation}
\lambda_z = 2 {{\Delta_p^2} \over \nu},
\label{lam}
\end{equation}
where $\Delta_p$ is the root mean square deviation of the distribution of
dipolar couplings
experienced by the muon spin in its interstitial stopping site. The rise of 
$\lambda_z$
below about 2\,K is thus indicative of a slowing down of the electronic
fluctuations as the spin correlations develop. Below the temperature of the
specific heat peak (0.24\,K), the shape of the muon depolarisation changes
drastically (see the 0.2\,K data in Fig.\ref{musr} ): it is no longer an 
exponential function of time, but it shows a rapid depolarisation within
about 0.2\,$\mu$s followed by a slow quasi-exponential decay. This shape
remains unchanged down to the lowest temperature of the experiment, 0.04\,K.
The first remarkable thing to notice is that no oscillatory signal is 
observed, although in this temperature range a hyperfine field is
observed in the M\"ossbauer spectra. This means that no LRO is present, in
agreement with the powder neutron diffraction data. The shape of the low 
temperature 
depolarisation can actually be accounted for by a dynamic Kubo-Toyabe decay
(dashed line in Fig.\ref{musr}), except for a small discrepancy around 
0.03\,$\mu$s: the dip present in the theoretical curve is blurred in the
experimental data. A dynamic Kubo-Toyabe lineshape describes the general case
of a muon spin coupled to an ensemble of electronic spins with fluctuation
frequency $\nu$, and with a root mean square isotropic coupling $\Delta$.
Its high frequency limit is an exponential function with a relaxation rate
given by expression (\ref{lam}). In order to reproduce 
correctly the low temperature depolarisation, a solution consists in 
introducing
a gaussian distribution of $\Delta$ values, which could be a means of taking
into account the spin correlations \cite{noakes}. A good fit is thereby 
obtained (solid line in Fig.\ref{musr} b), with a rather low fluctuation
frequency: $\nu \simeq 10^6$\,s$^{-1}$. The mean value
of the muon - 4f spin coupling, in units of field, is: $\Delta_{LT} /
\gamma_\mu \simeq 5.7$\,mT, where $\gamma_\mu = 851.6$\,Mrd s$^{-1}$T$^{-1}$
is the muon gyromagnetic ratio. Thus the presence, below 0.24\,K, of a $\mu$SR
signal consisting of a dynamic Kubo-Toyabe decay and of a hyperfine field
in the M\"ossbauer spectra, means that short range dynamic spin
correlations are present down to 0.04\,K. The measured $\mu$SR fluctuation 
frequency 
of $10^6$\,s$^{-1}$ is accordingly lower than the M\"ossbauer threshold
$\nu_l \simeq 10^8$\,s$^{-1}$.\par

Above the temperature of the specific heat peak, the electronic fluctuation
frequencies $\nu$ can be deduced from the $\lambda_z$ values through formula
(\ref{lam}) if the high temperature $\Delta_p$ value is known and also, in
principle, if the spin correlations are weak. Usually, in the paramagnetic
phase, fluctuating moments of 1\,$\mu_B$ (the size of the moments in
Yb$_2$Ti$_2$O$_7$) correspond to a mean dipolar field of
the order of 50\,mT. So the mean $\Delta_{LT}$ value of 5.7\,mT obtained
below 0.24\,K, where the spin correlations are well developed, is too small
and cannot be used at high temperature where correlations are weaker. So in
order to extract $\nu$ from the $\lambda_z$ data above 0.24\,K, we adjusted
$\Delta_p$ in expression (\ref{lam}) so that the frequency values match 
those measured by M\"ossbauer
spectroscopy (see Fig.\ref{freqt}). We obtained $\Delta_{HT} \simeq 80$\,mT,
which is the correct order of magnitude. In the paramagnetic 
phase, $\Delta_{HT}$ can
be calculated for a known crystal structure by assuming that the muon stops in 
a given interstitial site (and assuming also a dipolar only interaction with
the rare earth spin) \cite{dalya}. In the present case of an axially 
symmetric Kramers
doublet with g-tensor $\{g_z,g_\perp\}$, $\Delta_{HT}$ is given by the 
expression:
\begin{equation}
\Delta_{HT}^2 = { 1 \over {8\pi g_J^2}} [(g_z^2 - g_\perp^2) \Sigma_1 +
g_\perp^2 \Sigma_2 ],
\label{deltanis}
\end{equation}
where $\Sigma_1$ and $\Sigma_2$ are lattice sums whose expressions are
given in Ref.\cite{dalya}. There are three possible interstitial sites for 
the muon in the pyrochlore lattice for which: $\Delta_{HT}/\gamma_\mu \simeq 
150$\,mT. This value is larger by a factor of 2 than
that obtained by scaling the $\mu$SR data to the M\"ossbauer data. This is
an acceptable agreement, if one keeps in mind that expression (\ref{deltanis})
is valid for uncorrelated moments only. In fact, at least below 2\,K, the
specific heat data and the neutron diffraction data on a single crystal (see
below) show that spin correlations are present. Therefore, the muon spin
is likely to interact with a well defined rare earth moment, of
modulus 1.15\,$\mu_B$ (the low temperature value), within correlated clusters,
rather than with a paramagnetic Kramers doublet. With this assumption, the
standard calculation of $\Delta_{HT}$ due to randomly oriented
moments yields $\Delta_{HT}/\gamma_\mu \simeq 80$\,mT for the three
sites, which is exactly the value derived above.\par

The overall thermal variation of the Yb spin fluctuation frequency is 
represented in Fig.\ref{freqt} and constitutes the main result of our study
in this compound: the fluctuation frequency undergoes a first order abrupt
drop at the temperature of the specific heat anomaly (0.24\,K), falling
from the 10$^4$-10$^5$\,MHz range to 1\,MHz. So the transition involves the
time domain and is reminiscent
of the liquid-gas transition, which is first order and involves an abrupt 
drop of the mean collision frequency between atoms as one enters the liquid
phase. 

The low temperature phase in Yb$_2$Ti$_2$O$_7$ is a ``spin-liquid''
phase, as predicted by Villain \cite{villain} for spins in a lattice of
corner sharing tetrahedra, where spin dynamics is present down to the lowest 
temperature. However,
this state is not reached ``smoothly'' by a continuous decrease of the
fluctuation frequency, but by a first order transition in the fluctuation
frequency. At this stage, it is not known whether the transition is 
accompanied by some change in the spatial spin correlations. The answer to
this question comes from a neutron diffraction study of a single crystal,
to be described next.

\begin{figure} [!ht]
\begin{center}
\includegraphics[width=0.7\textwidth]{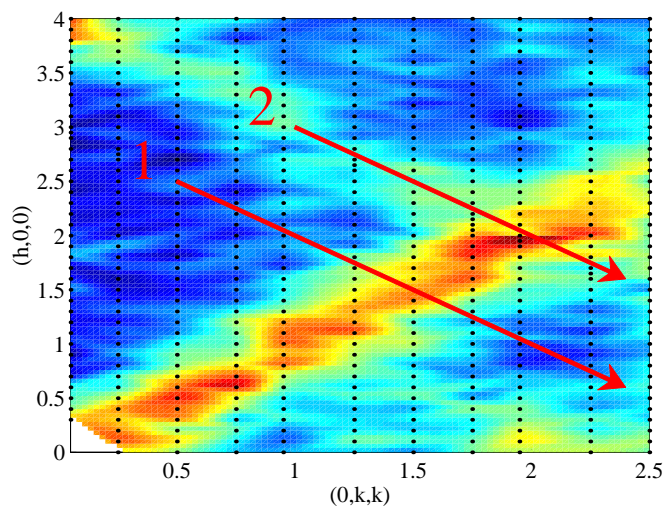}
\end{center}
\caption{\sl [color] Map of the magnetic neutron scattering in a 
single crystal of Yb$_2$Ti$_2$O$_7$ in the (110) plane of the reciprocal 
space at 1.4\,K, showing the ``diffraction rod'' along [111]. The red lines
labelled 1 and 2 are the directions along which Q-scans were performed.}
\label{single}
\end{figure}

\begin{figure} [!ht]
\includegraphics[width=0.5\textwidth]{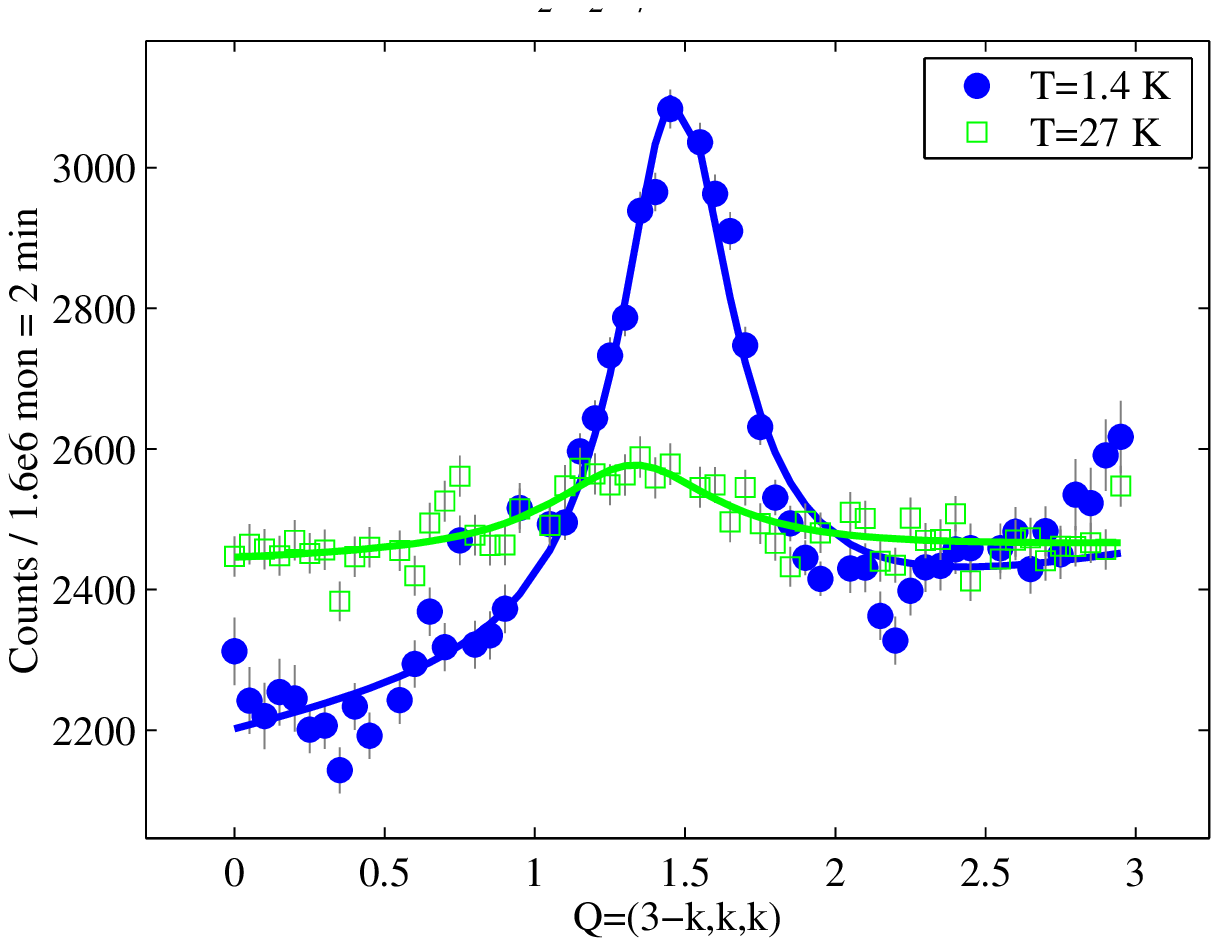}
\includegraphics[width=0.5\textwidth]{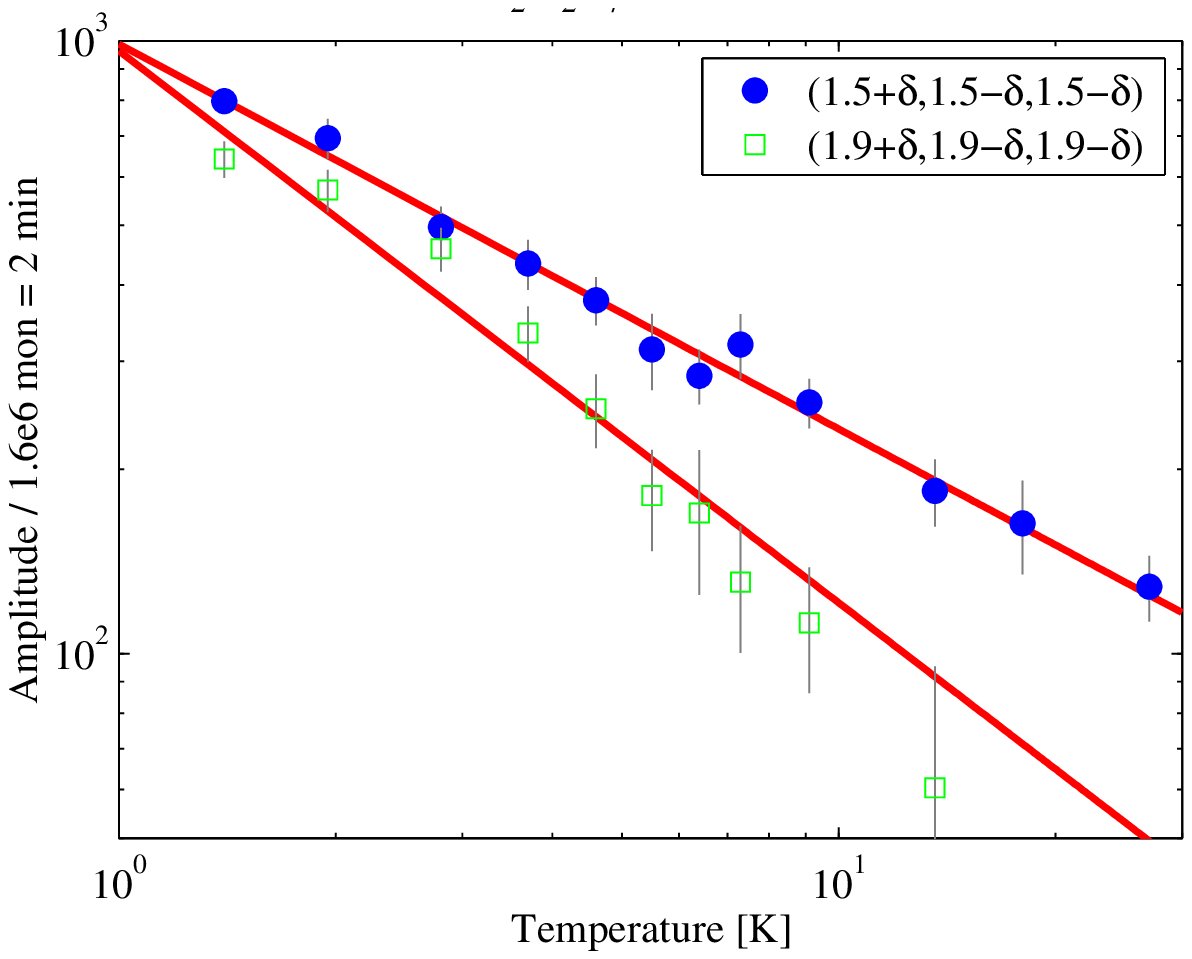}
\caption {\sl [color] \textbf{Left}: Q-scan across line 1 of the map in
Fig.\protect\ref{single} at 1.4 and 27\,K, and \textbf{Right}: thermal
variations of the amplitudes of the peaks
near (1.5,1.5,1.5) across line 1 and near (1.9,1.9,1.9) across line 2 of the 
map in Fig.\protect\ref{single}.}
\label{ampl}
\end{figure}

The neutron diffraction experiments on a single crystal of Yb$_2$Ti$_2$O$_7$
were performed at the Institut Laue Langevin (Grenoble, France), in the
temperature range 0.04\,K - 30\,K. Like the previous neutron diffraction
measurements on a powder sample, no magnetic Bragg peaks were discovered
as temperature decreases from 4.2\,K to 0.04\,K. But mapping of the magnetic
elastic scattering in (110) planes reveals the presence of ``diffraction 
rods'' along the [111] direction in reciprocal space (see Fig.\ref{single}). 
This is indicative of the presence of
bidimensional antiferromagnetic spatial correlations, probably within planes
perpendicular to [111]. No anomaly is found at 
0.24\,K: the intensities of the peaks obtained from Q-scans across the rod
(see Fig.\ref{ampl} left for the scan along line 1) increase
steadily on cooling from about 25\,K down to 0.04\,K (see Fig.\ref{ampl} 
right, where the data below 1.4\,K are not shown).
The width of the peaks in Q-space does not vary with temperature and 
corresponds to a correlation length $\xi \simeq 4$\,nm (about 4 times the
parameter of the unit cell, which contains 16 Yb ions). The short range
correlations therefore involve a few hundred Yb ions.
 
The magnetic scattering intensity near \textbf{Q}=(1.5,1.5,1.5) starts to
develop near 25\,K, whereas that near \textbf{Q}=(1.9,1.9,1.9) starts to grow
at a lower temperature (about 13\,K). Both peaks however reach half their 
maximum
intensity around 2-3\,K. This suggests that these spin correlations are
driven by the exchange evidenced in the specific heat bump, which has the
same energy scale of 2-3\,K. The short range bidimensional spin correlations 
in Yb$_2$Ti$_2$O$_7$ thus build up monotonically as temperature decreases; the
range of these correlations, about 4\,nm, is temperature independent and in
particular shows no anomaly at 0.24\,K. This reinforces the picture sketched
above, of a transition involving the frequency domain only.\par

\section{Gd$_2$Sn$_2$O$_7$}

\begin{figure} [!ht]
\includegraphics[width=0.47\textwidth]{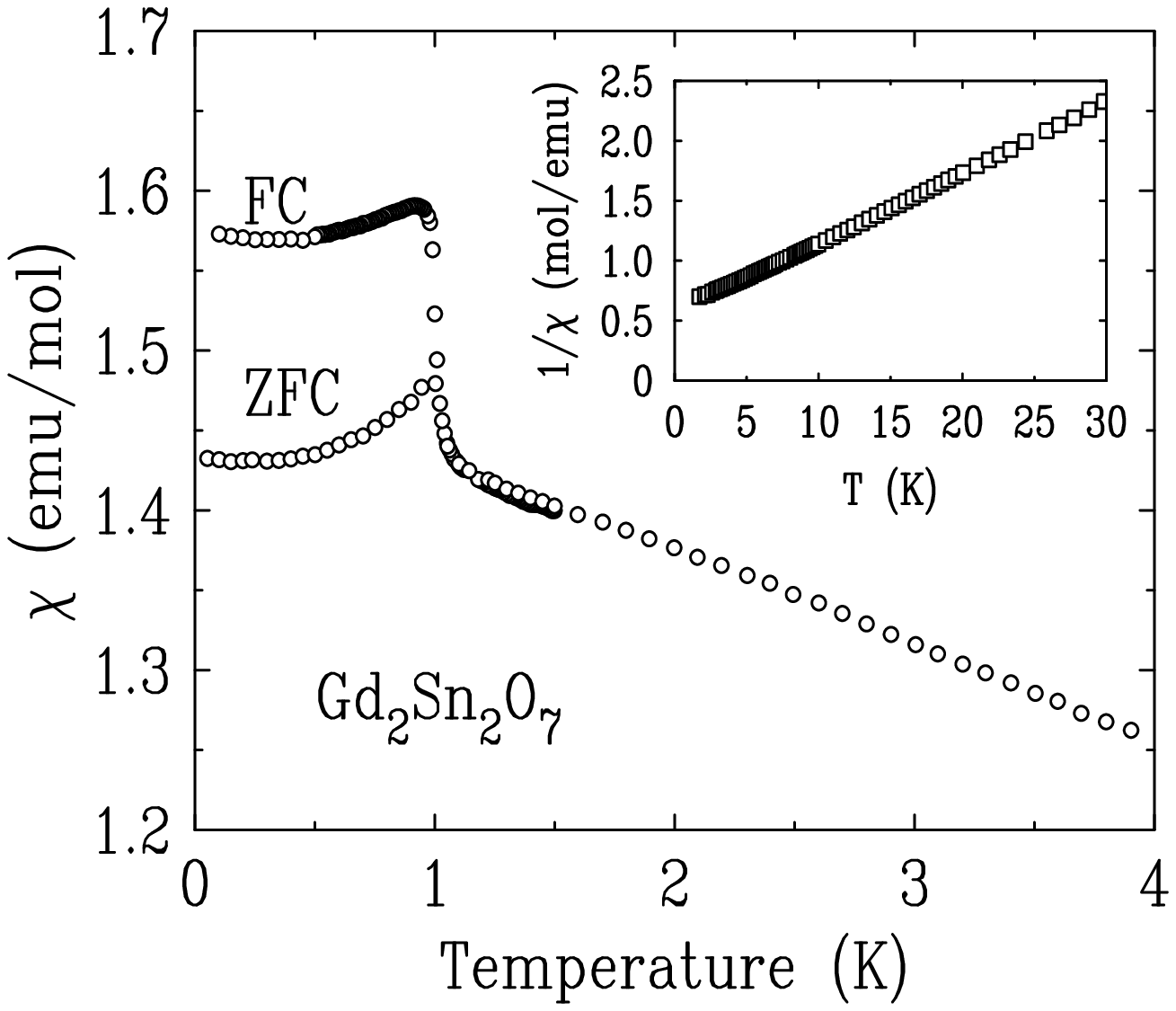}
\hspace{0.7 cm}
\includegraphics[width=0.47\textwidth]{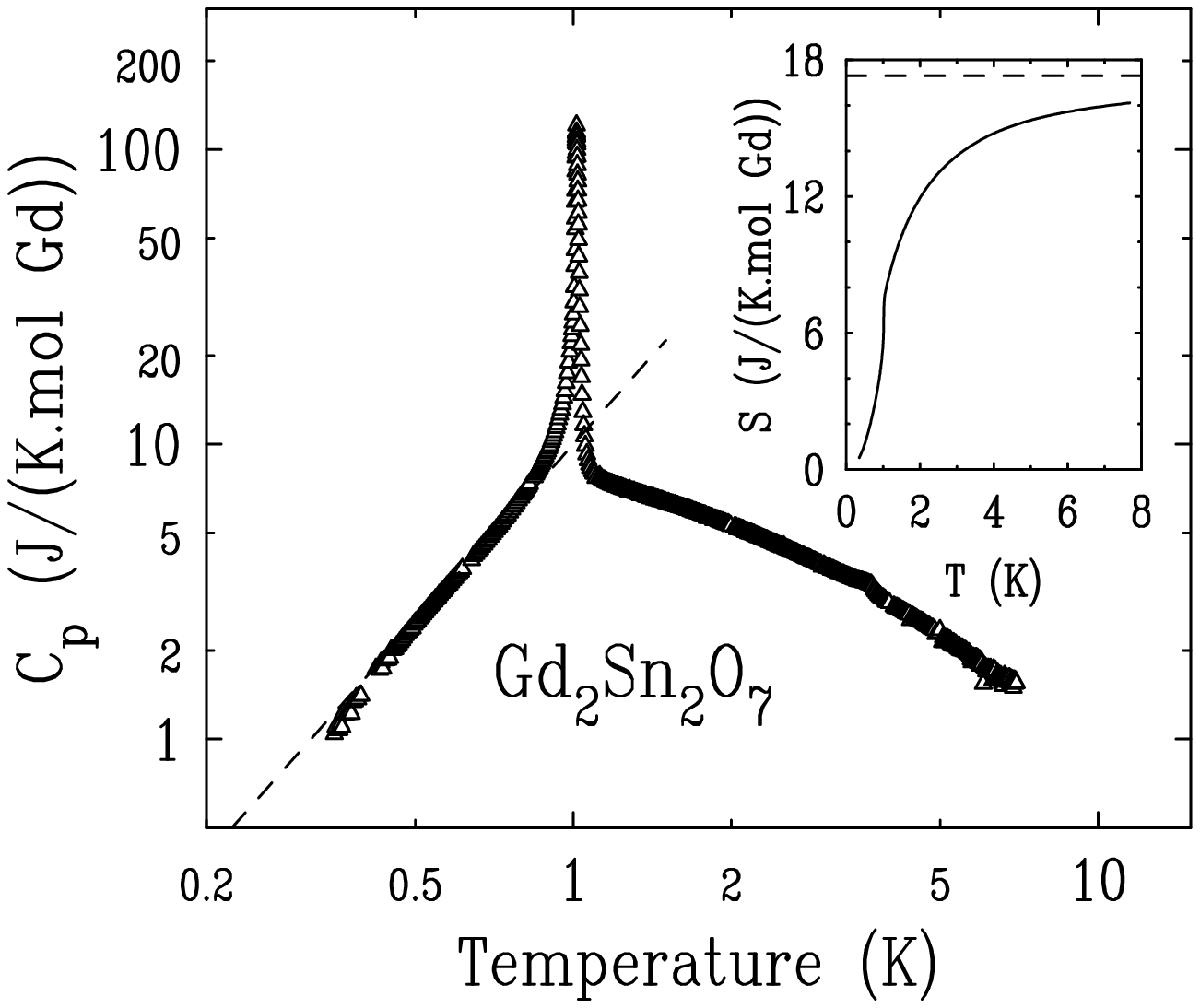}
\caption {\sl In Gd$_2$Sn$_2$O$_7$: \textbf{Left}: magnetic susceptibility 
measured with 1\,mT, with the Field Cooled (FC) and Zero Field Cooled (ZFC) 
branches; inset: inverse susceptibility below 30\,K, and \textbf{Right}: 
thermal variation of the
specific heat; the dashed line is a $T^2$ law; inset: thermal variation of the
entropy.}
\label{cpsusc}
\end{figure}

As shown in the inset of Fig.\ref{cpsusc} left, the inverse susceptibility in 
Gd$_2$Sn$_2$O$_7$ follows a Curie-Weiss law with a paramagnetic Curie 
temperature $\theta_p \simeq -10$\,K, indicative of an antiferromagnetic 
exchange interaction. As the Gd$^{3+}$ ion is isotropic, this compound can be 
therefore expected to be a good realisation of an AF Heisenberg frustrated 
system, with no N\'eel order down to $T=0$. However, both the magnetic
susceptibility and the specific heat, shown in Fig.\ref{cpsusc} right and left
respectively, evidence an anomaly at 1\,K. The susceptibility $\chi(T)$
presents furthermore a sizeable irreversibility between the FC and ZFC 
branches below 1\,K. The anomaly of the specific heat $C_p(T)$ at 
1\,K reaches the very large value of 
120\,JK$^{-1}$mol.Gd$^{-1}$, whereas the expected jump at $T_N$ for a
second order magnetic transition is: $\Delta C_p = {5 \over 2} R {{S(S+1)}
\over {S(S+1)+0.5}}$ \cite{stanley}, i.e. 20.4\,JK$^{-1}$mol.Gd$^{-1}$ for
Gd$^{3+}$ with $S$=7/2.  

This high $\Delta C_p$ value is indicative of a first order transition (the
specific heat diverges in principle in this case), which
will be confirmed by the $^{155}$Gd M\"ossbauer measurements to be described 
below.
Below 1\,K, the specific heat varies as $T^2$. Classically, the specific
heat due to the thermal excitation of magnons in a 3-dimensional Heisenberg AF
varies as $T^3$, and as $T^2$ in the bidimensional case. So the observed
quadratic variation in Gd$_2$Sn$_2$O$_7$ suggests a bidimensional AF 
structure.  
The inset of Fig.\ref{cpsusc} left shows the magnetic entropy variation as 
temperature increases; at the transition, only 40\% of the total entropy
$R\ln 8=17.3$\,JK$^{-1}$mol.Gd$^{-1}$ has been released, and the full
paramagnetic degrees of freedom are recovered only at 8-10\,K. As in the case
of Yb$_2$Ti$_2$O$_7$, this is due to the presence of short range order 
developing well above the transition temperature and it evidences the presence
of frustration.\par
\begin{figure} [!ht]
\includegraphics[width=0.3\textwidth]{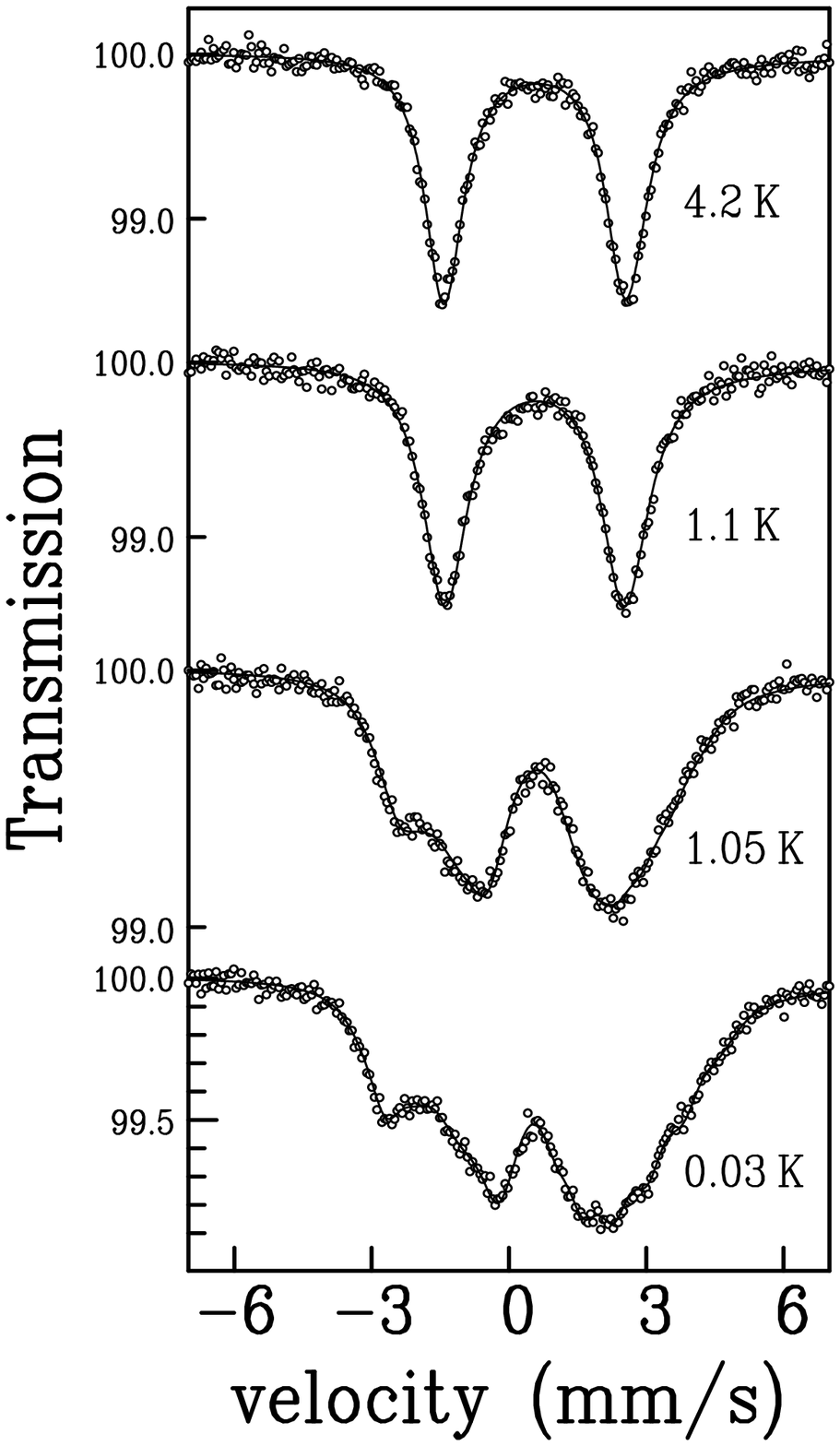}
\hspace{0.5 cm}
\includegraphics[width=0.6\textwidth]{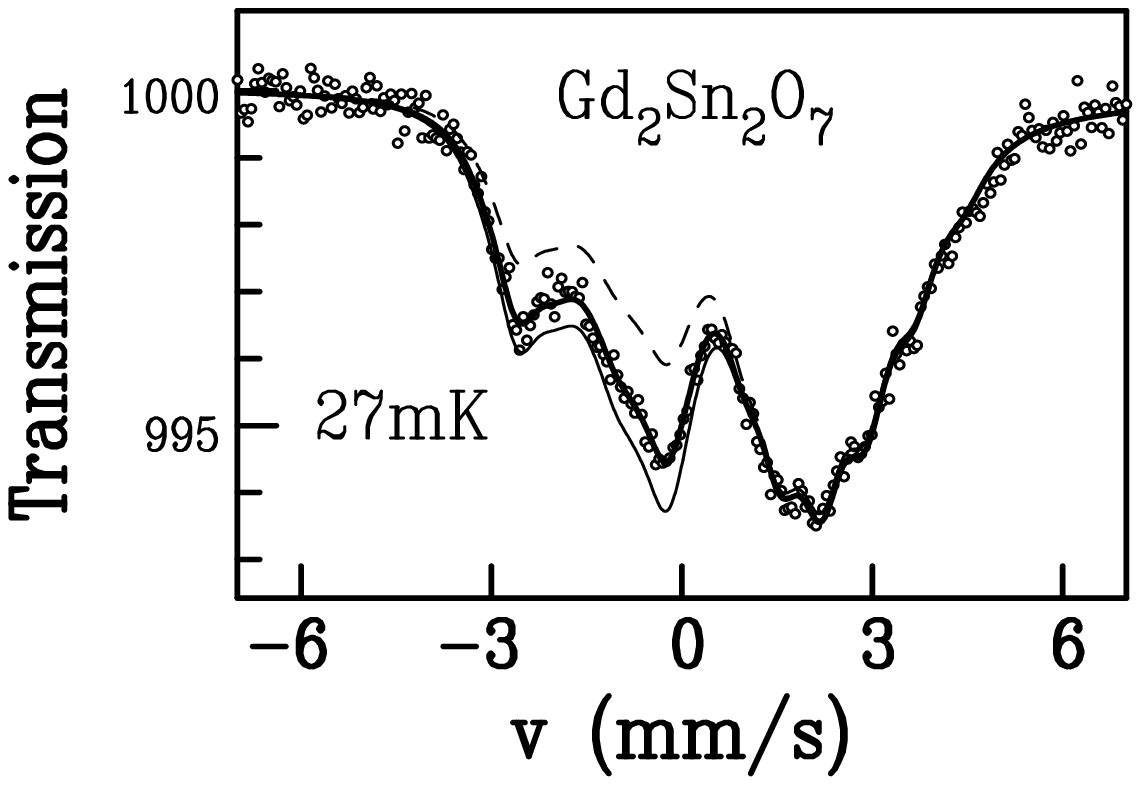}
\caption {\sl $^{155}$Gd M\"ossbauer spectra in Gd$_2$Sn$_2$O$_7$: 
\textbf{Left}: at selected temperatures between 0.027\,K and 4.2\,K, and 
\textbf{Right}: at 0.027\,K; the lines are theoretical curves for various
values of the hyperfine temperature $T_{hf}$: dashed line, $T_{hf}$=0.027\,K,
thick solid line, which fits the data well, $T_{hf}$=0.090\,K, and thin solid 
line: high temperature limit, $T_{hf} > 0.2$\,K.}
\label{spegd}
\end{figure}

Selected M\"ossbauer absorption spectra on the isotope $^{155}$Gd ($I_g$=3/2,
$I_e$=5/2, $E_0$=86.5\,keV) in Gd$_2$Sn$_2$O$_7$ are shown in Fig.\ref{spegd}.
A clear change in the lineshape is observed between 1.05 and 1.1\,K (see
Fig.\ref{spegd} left): at 1.1\,K and above, the spectrum is a pure quadrupolar
hyperfine pattern characteristic of the paramagnetic phase, while at 1.05\,K
a magnetic hyperfine field has appeared, evidencing short or long range 
magnetic order. As the temperature is further decreased
down to 0.027\,K, the hyperfine field increases slightly to reach a saturated
value of 30\,T. This behaviour is the hallmark of a first order transition,
as also inferred from the height of the specific heat peak; in 
Gd$_2$Sn$_2$O$_7$
however, the coexistence region between the two phases could not be detected
like in Yb$_2$Ti$_2$O$_7$. The fits furthermore show that each Gd$^{3+}$ 
moment is perpendicular to the [111] local symmetry axis. 

The question arises here also as to whether there is LRO below 1\,K, or 
whether the
correlated moments fluctuate with a frequency lower than $\nu_l \simeq 3 
\times 10^7$\,s$^{-1}$, which is the lower limit of the ``M\"ossbauer window''
for $^{155}$Gd. The answer will be provided by the $\mu$SR measurements, but 
before we turn to their description, we will examine the problem posed by the 
lowest temperature M\"ossbauer spectrum.\par
The spectrum at 0.027\,K (see Fig.\ref{spegd} right) indeed deserves 
particular attention, because the temperature which can be measured from the 
spectrum ($\simeq 0.090$\,K, referred to hereafter as the hyperfine 
temperature $T_{hf}$), is much higher than that of the sample (given by the 
thermometer): $\simeq 0.03$\,K. The possibility of measuring the absolute 
temperature from the relative line intensities of the 
M\"ossbauer spectrum has been since long recognised \cite{shenoy}. The 
conditions are that the spin $I_g$ of the ground nuclear state is non-zero,
and that
the hyperfine splittings are of the same order of magnitude as $k_BT$.
Then each resonant transition has an intensity which is weighed by the 
population of the ground hyperfine level it arises from. The temperature of
the hyperfine levels (in principle equal to the lattice temperature) can then
be obtained directly from the spectrum. This procedure is impossible for 
$^{170}$Yb, which has $I_g$=0, but should work for $^{155}$Gd, which has
$I_g$=3/2. In Gd$_2$Sn$_2$O$_7$, the combined quadrupolar and magnetic 
hyperfine interactions yield four ground hyperfine levels with energies 0,
0.05, 12.1 and 15.9\,mK, i.e. two quasi-degenerate doublets separated by a 
mean hyperfine splitting $\Delta_{hf} \simeq 14$\,mK. Then the hyperfine 
temperature can be measured if it is below about
0.15\,K. On Fig.\ref{spegd} right are shown, together with the experimental 
spectrum at 27\,mK, the theoretical
spectra expected for $T_{hf} = 27$\,mK (dashed line), for $T_{hf} > 0.2$\,K
(thin solid line) and for $T_{hf} = 90$\,mK (thick solid line). This last line
goes perfectly well through the data points, and it is then clear that the 
hyperfine levels are ``hotter''
than the lattice, i.e. they are out of thermal equilibrium. 

\begin{figure} [!ht]
\begin{center}
\includegraphics[width=0.6\textwidth]{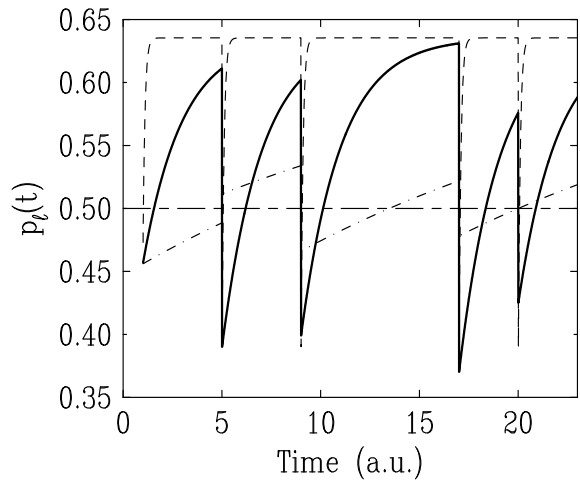}
\end{center}
\caption {\sl Temporal evolution of the ground state population $p_\ell(t)$
of a (nuclear) spin
1/2 doublet, as the (hyperfine) field reverses randomly in time, at 
arbitrarily chosen instants 5, 9, 17 and 20. The ratio
of the doublet splitting to $k_BT$ is chosen to be about 2, so that the
Boltzmann population of the ground state is 0.63. Dashed line:
$T_1 \ll \tau$: thermalisation is effective; dash-dotted line: $T_1 \gg \tau$:
thermalisation is impossible, and solid line: $T_1 \sim \tau$.}   
\label{p_pB}
\end{figure}

The interpretation
we give for this phenomenon is the following \cite{bertin}: the hyperfine
levels have no time to reach thermal equilibrium, with a time constant $T_1$,
because the hyperfine field reverses with a time constant $\tau$ shorter than
or of the same magnitude as $T_1$. This is illustrated in Fig.\ref{p_pB},
which depicts the temporal evolution of the population $p_\ell$ of the ground 
state of a fictitious spin 1/2 (modelling the hyperfine levels) as a function 
of the ratio $T_1 / \tau$. The steady state population (i.e. time averaged)
$\langle p_\ell \rangle$ is the Boltzmann population $p_\ell^B$ if hyperfine
relaxation occurs very rapidly with respect to $\tau$ (dashed line in 
Fig.\ref{p_pB}), it is 0.5 (equipopulation) in the reverse case (dash-dotted 
line in Fig.\ref{p_pB}), and it should be a function of the ratio
$\mu = T_1/\tau$ if the latter is of the order unity (solid line in 
Fig.\ref{p_pB}). In order to obtain this function, we devised a model of
a spin 1/2 submitted to a magnetic field of constant magnitude, but which 
undergoes ``flips'' at random
instants of time. The model provided an analytical solution for the
steady state population of the ground level:
\begin{equation}
\langle p_\ell \rangle = {1 \over 2} (1+{1\over {1+2\mu}} \tanh{\Delta \over
{2k_BT}}),
\label{pell}
\end{equation}
where $\Delta$ is the Zeeman splitting of the spin 1/2 doublet. Assigning
to the doublet an effective (in our case hyperfine) temperature (this is
always possible for a spin 1/2 system), we obtain: $T_{hf} \simeq T (1+2\mu)$
in the range $k_BT > 2\Delta$ which is the temperature range of our 
experiments. Applying this relation to the case of Gd$_2$Sn$_2$O$_7$ yields
$\mu \simeq 0.8$. \par
The fact that the hyperfine populations are out of equilibrium at 0.027\,K 
reveals therefore the presence of two forms of dynamics of the Gd$^{3+}$
electronic spin. The first one consists in  spin flips between one (or 
more) direction(s) which, as the Gd$^{3+}$ moment is proportional to the 
hyperfine field, correspond to flips of the hyperfine field sensed by the
$^{155}$Gd nuclear spin. Besides, the nuclear (hyperfine) relaxation, at such 
low temperature, cannot be due to phonons. The only plausible mechanism is 
nuclear relaxation driven by coupling to electronic spin-waves \cite{beeman}, 
which are thus the second form of spin dynamics evidenced at 0.027\,K in 
Gd$_2$Sn$_2$O$_7$. The presence of spin-waves at very low temperature
is also demonstrated by the $\mu$SR experiments, to be
described below. The probe in that case is not the nuclear spin of the rare
earth atom itself, but the spin of positive muon lying at an interstitial 
site; the relaxation mechanisms are however similar in both cases.\par

\begin{figure} [!ht]
\includegraphics[width=0.5\textwidth]{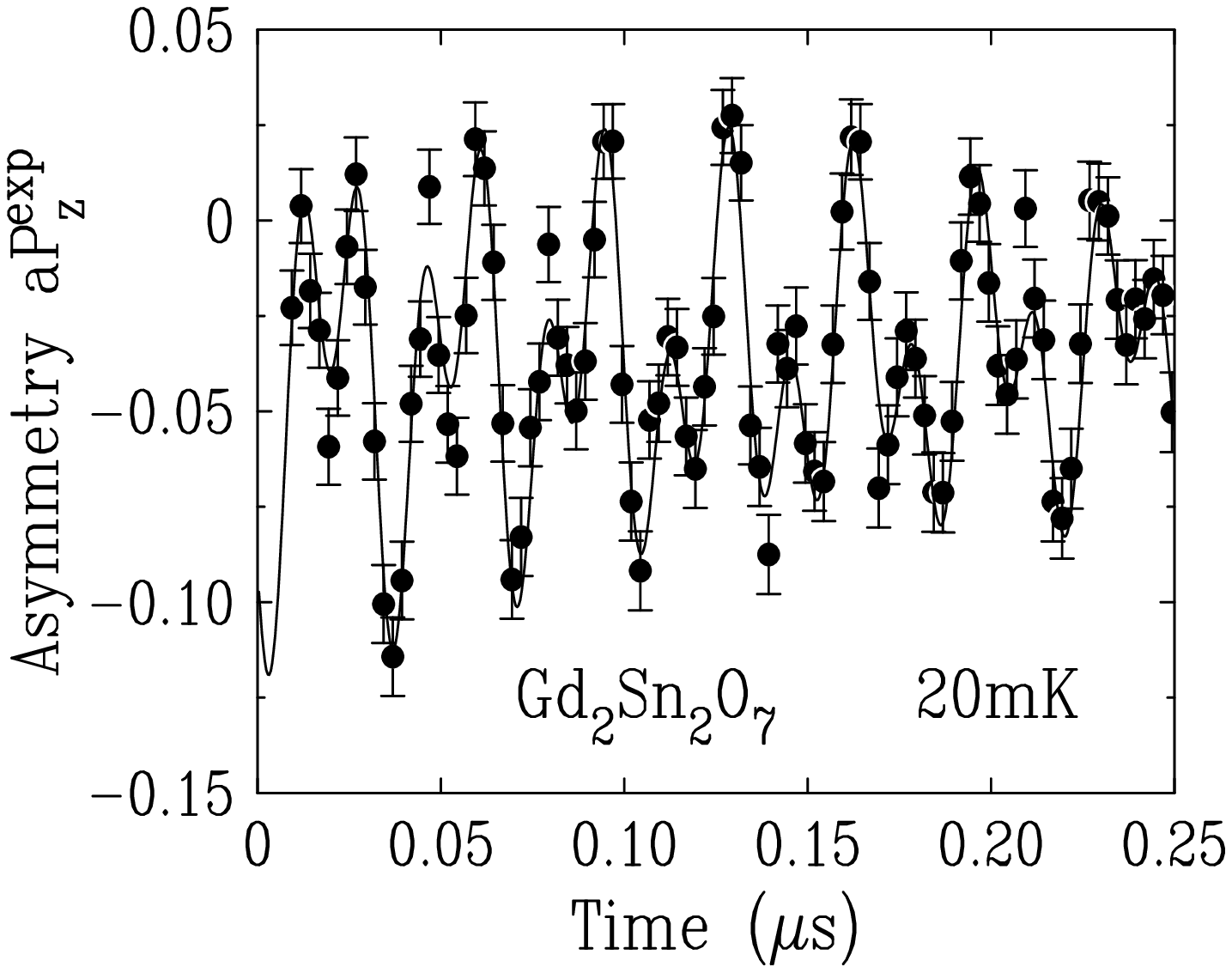}
\includegraphics[width=0.5\textwidth]{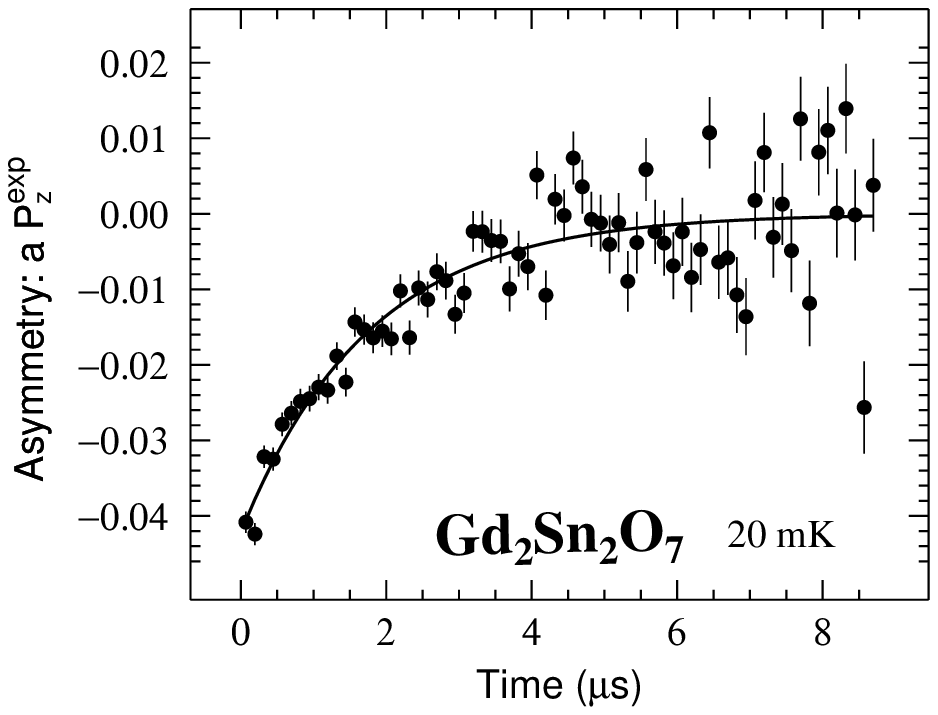}
\caption {\sl Time variation of the $\mu$SR depolarisation in 
Gd$_2$Sn$_2$O$_7$ at 20\,mK, measured at PSI:
\textbf{Left}: short time part showing the rapidly oscillating components, 
and \textbf{Right}: long time part with a 100 channel binning in order to
integrate out the oscillations.}
\label{musr20mk}
\end{figure}

The $\mu$SR experiments were performed in zero magnetic field between 20\,mK 
and 100\,K at PSI. Above 1\,K, the decay has an exponential form,
and the relaxation rate $\lambda_z$ has a temperature independent value
of 2\,MHz. Below 1\,K, the decay signal changes and presents an oscillating
component, indicative of magnetic LRO. The oscillations are clearly visible
only at very short times (see Fig.\ref{musr20mk} left).

An exponential damping component is also present, which is best visible 
when the counting channels are grouped into batches of 100 (``binning'' of
100 channels), as shown in Fig.\ref{musr20mk} right where the line is a fit to
an exponential function. Usually, in the magnetically ordered phase,
the depolarisation in a polycrystalline sample is well described by the 
following expression (assuming
a single stopping site for the muon) \cite{schenk}:
\begin{equation}
aP_z(t) = {a \over 3} [\exp(-\lambda_z t) + 2 \exp(-\lambda_\perp t)
\cos\omega_\mu t].
\label{muord}
\end{equation}
This is equivalent to a depolarisation function where one third of the
impinging muons have their spin parallel to the internal field (first term) 
and two thirds
have their spin precessing in the internal (dipolar) field (second term). 
In expression (\ref{muord}), $a$ is the amplitude (or asymmetry) of the 
depolarisation,
which in principle does not change when the temperature is lowered through the
transition; $\lambda_z$ is the longitudinal relaxation rate, due to the
coupling of the muon spin with the electronic degrees of freedom 
(in the LRO phase, the spin-waves) \cite{ya}; $\omega_\mu = \gamma_\mu B_i$ 
is the
Larmor frequency of the muon spin in the internal field \textbf{B}$_i$, 
arising from the spontaneous electronic moment, and 
$\lambda_\perp$ is a damping factor which is generally taken to be due both
to a distribution in $B_i$ values and to fluctuations of \textbf{B}$_i$. As 
discussed above,
the depolarisation in Gd$_2$Sn$_2$O$_7$ below 1\,K is described by
expression (\ref{muord}) in a first approximation. However, some features are
not fully understood at the present time: the asymmetry $a$ below 1\,K is 
half the value
above 1\,K ($a \simeq 0.2$); there are two oscillating components of equal 
weights, with dipolar fields of 206 and 441\,mT, each with a sizeable phase 
shift, respectively 24 and 69$^\circ$. A particular magnetic structure could 
be at the origin of these peculiarities. We will only mention here that the
two dipolar fields are temperature independent between 20\,mK and 0.7\,K,
the highest temperature where the oscillations are clearly visible. The
dipolar fields at the muon sites are due to the spontaneous Gd$^{3+}$ moments,
and their lack of thermal variation is in line with the M\"ossbauer 
spectroscopy finding of an essentially temperature independent moment value, 
i.e. of a
first order transition. Their order of magnitude (200-400\,mT) is in
agreement with the rough rule stating that a moment of 1\,$\mu_B$ gives rise 
to a dipolar field of 50\,mT, as the Gd$^{3+}$ saturated moment is 7\,$\mu_B$.
\par

\begin{figure} [!ht]
\begin{center}
\includegraphics[width=0.6\textwidth]{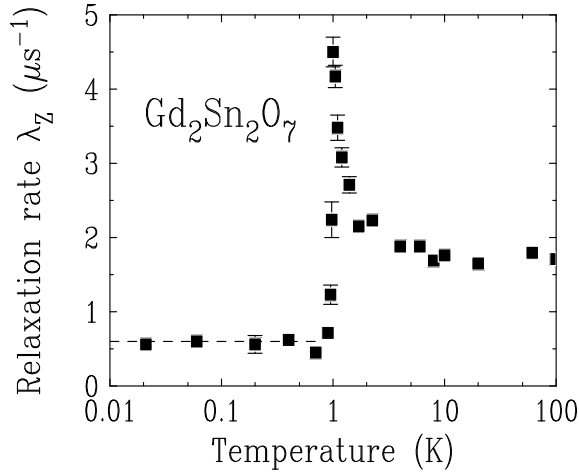}
\end{center}
\caption {\sl Thermal variation of the muon longitudinal relaxation rate 
$\lambda_z$ in Gd$_2$Sn$_2$O$_7$. Below 1\,K, $\lambda_z$ is defined by
Eqn.(\ref{muord}); above 1\,K, it is the relaxation rate of the observed
exponential decay (the data for $T \ge 4$\,K were obtained with a longitudinal
magnetic field of 10\,mT). The dashed line at 0.6\,MHz underlines the constant
relaxation rate below 1\,K.}
\label{lamzt}
\end{figure}

The thermal variation of the relaxation rate $\lambda_z$ is shown in
Fig.\ref{lamzt}. Below 1\,K, it reflects the muon spin lattice relaxation
through the coupling with spin-waves, and it is defined by Eqn.(\ref{muord});
above 1\,K, it reflects the fast relaxation due to coupling to the 
short range correlated, and eventually paramagnetic, Gd$^{3+}$ spins.  

The transition at 1\,K is marked by a sharp anomaly in $\lambda_z(T)$. The
remarkable feature is that $\lambda_z$ below 1\,K shows no thermal dependence,
remaining at the constant value 0.6\,MHz down to 20\,mK. This is anomalous,
as relaxation of the muon spin by spin-waves should vanish at $T=0$. For
example, it was shown that, for a Heisenberg ferromagnet, $\lambda_z$ varies
as $T^2$ to a good approximation if one considers two-magnon Raman processes,
which are the most likely to occur for relaxation between the quasi-degenerate
spin states of the muon \cite{ya}. An analogous calculation for a Heisenberg 
bidimensional AF, which could apply to Gd$_2$Sn$_2$O$_7$ as suggested by the 
specific heat data, yields a $T^3$ variation. None of these thermal laws
matches with experiment. The explanation for the temperature independence of
$\lambda_z$ must probably be looked for in the unusual spin-wave spectrum
of Gd$_2$Sn$_2$O$_7$, where soft spin-wave modes with a finite density of 
states at zero energy could exist. A saturation of $\lambda_z$ below about
2\,K has also been observed in another pyrochlore material, Tb$_2$Ti$_2$O$_7$
\cite{gardner}. But this compound shows no phase transition to an LRO phase
at low temperature and remains a ``correlated paramagnet'' down to $T=0$.
\par
  
\section{Conclusions}

Both studied pyrochlore materials, Yb$_2$Ti$_2$O$_7$ and Gd$_2$Sn$_2$O$_7$,
present a low temperature state where spin fluctuations persist as $T \to 0$.
\par
In Yb$_2$Ti$_2$O$_7$, this state seems to be a spin-liquid phase where short
range dynamic spin correlations are present. Such a phase is expected for 
geometrically
frustrated isotropic systems with antiferromagnetic interactions. However, 
it is not reached by a smooth decrease of the fluctuation frequency as the 
temperature decreases, but through a first order transition (at 0.24\,K) 
where the spin fluctuation frequency drops from the
GHz to the MHz range. Below 0.24\,K, this frequency could be measured directly
by the $\mu$SR experiments and it remains constant down to 40\,mK. No anomaly
could be detected at the transition for the spatial spin correlations: they
build up from 10-20\,K and remain short range, with a correlation length of 
4\,nm, down to the lowest temperature.\par
In Gd$_2$Sn$_2$O$_7$, the low temperature phase has long range magnetic order
and it is reached through a standard first order transition. However, spin
dynamics is present down to 30\,mK: it was detected indirectly
by $^{155}$Gd M\"ossbauer spectroscopy through the finding of nuclear levels
which are out of thermal equilibrium. This requires the presence of both
electronic spin flips and of nuclear relaxation by coupling to spin waves,
with similar characteristic time scales. The presence of spin waves down to
20\,mK was also evidenced in the $\mu$SR measurements by the non-vanishing 
relaxation rate of the $\mu^+$ spin. \par
These findings illustrate the unconventional dynamics occurring in the
frustrated pyrochlores at very low temperature and the diversity of
routes leading to a ground state where spin fluctuations persist as $T \to 0$.

\bigskip
\textbf{Acknowledgements}: We thank A. Forget from SPEC-Saclay for preparing 
the powder samples and G. Dhalenne, from LPCES-Orsay, for growing the single
crystal Yb$_2$Ti$_2$O$_7$ sample. The neutron scattering experiments on this
single crystal were performed at the D23 diffractometer at the Institut
Laue-Langevin, which is a Collaborating Research Group (CRG) instrument
operated by the CEA.

\end{document}